\documentclass[doublecol]{epl2}

\usepackage{graphicx}

\title{How dense can one pack spheres of arbitrary size distribution?}

\author{Saulo D. S. Reis\inst{1} \and Nuno A. M. Ara\'{u}jo\inst{2}
\and Jos\'e S. Andrade Jr.\inst{1,2} \and Hans J. Herrmann\inst{1,2}}

\institute{
  \inst{1} Departamento de F\'{\i}sica, Universidade Federal
do Cear\'a, 60451-970 Fortaleza, Cear\'a, Brazil\\
  \inst{2} Computational Physics for Engineering Materials, IfB,
ETH Z\"{u}rich, Schafmattstr. 6, CH-8093 Z\"{u}rich, Switzerland
}


\abstract{
We present the first systematic algorithm to estimate
the maximum packing density of spheres when the grain sizes are drawn
from an arbitrary size distribution.  With an Apollonian filling rule,
we implement our technique for disks in 2d and spheres in 3d.  As
expected, the densest packing is achieved with power-law size
distributions.  We also test the method on homogeneous and on empirical
real distributions, and we propose a scheme to obtain experimentally
accessible distributions of grain sizes with low porosity.  Our method
should be helpful in the development of ultra-strong ceramics and high
performance concrete.  
}

\begin{document}

\maketitle

\begin{figure}[b]
  \includegraphics[width=0.9\columnwidth]{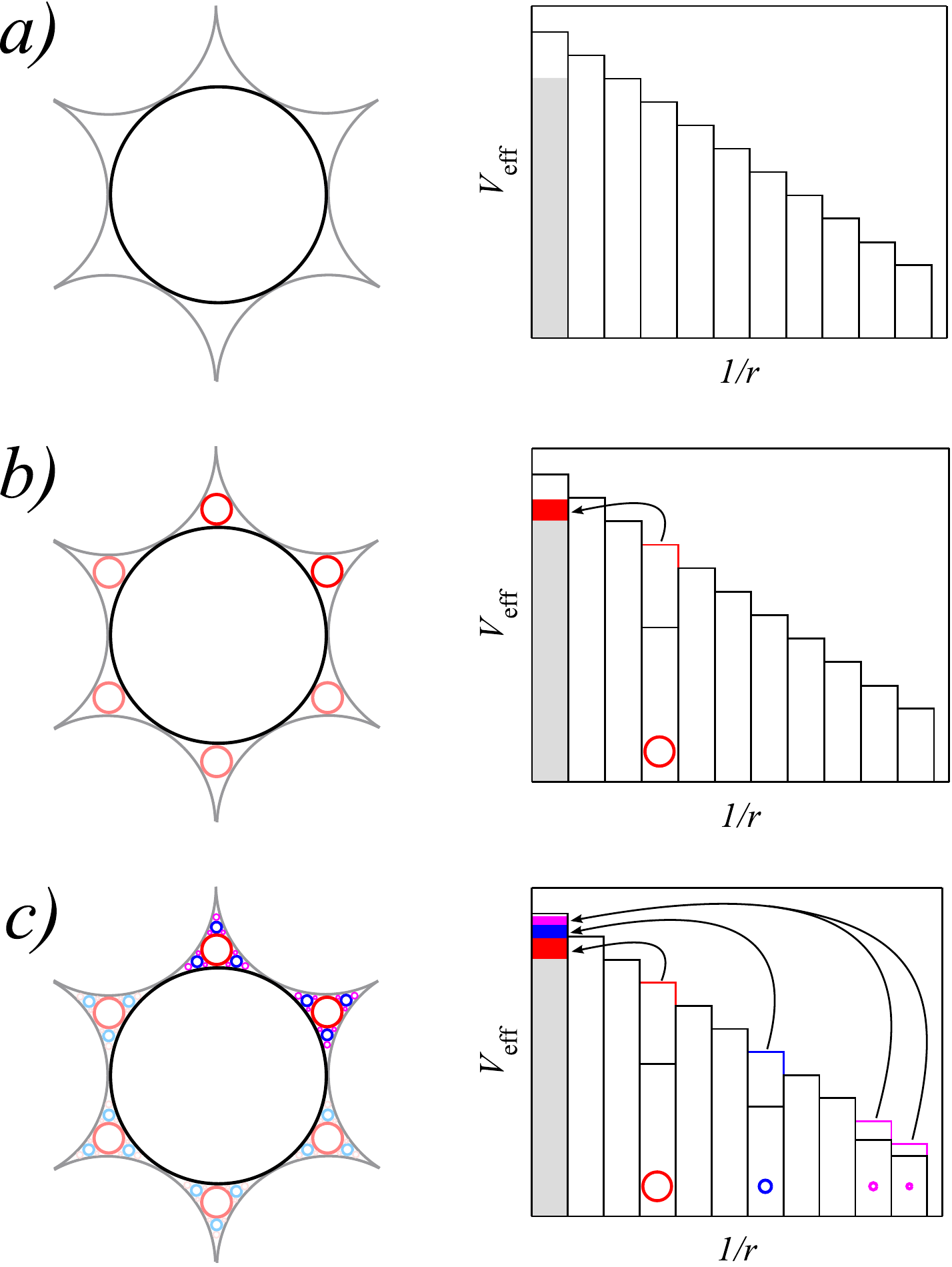}
  \caption{
    Schematic description of the algorithm in 2d. 
    a) Using the {\it hcp} as the initial packing of a particular bin, the algorithm fills the empty spaces using the positive solution of Eq.~(\ref{eq::soddynD}). 
    b) By symmetry, for each largest disk, two voids are to be filled.
    c) In an arbitrary distribution of effective volumes (right-hand side), we remove the amount of volume required according to the procedure described in the text. 
       In the beginning all bins are filled with the {\it hcp}.
       Starting from the bin of the largest grains, voids are filled with grains from the further bins: forth bin (red) for the first generation, seventh bin (blue) for the second, and tenth and eleventh bins (magenta) for the third.
       The regions shown in the first bin represent the final contribution of each grain size to the effective volume, where white stands for the empty space.
       Proportion between sizes has been exaggerated for clarity.
    \label{fig::scheme}
    }
\end{figure}

  High strength ceramics and ultra-high performance concrete (UHPC)~\cite{Fehling08} require minimizing the porosity out of a compacted either reactive or sintered powder.
  Much research effort has been invested in the past to optimize this procedure and the most important factor turned out to lie in the adequate choice of the size distribution of the constituents.
  In fact, the broader this distribution, the smaller result the remaining voids.
  So, mixtures of very different grains with up to four orders of magnitude in grain size are typically used for UHPC.
  But what is the ideal partitioning?
  Which combined grain-size distribution would yield the most compact mixture?
  The key to answer this fundamental question posed by the practitioners is to be able to estimate the maximum density a given distribution can provide.
  This is precisely the aim of this Letter.

  The maximal filling density has been studied for many specific types of packings~\cite{Aste08} starting with the work of Fuller~\cite{Fuller1907}.
  Besides the simple monodisperse case also exact results for some symmetric cases are known~\cite{Torquato09,Jiao08}.
  Various models have been proposed to deal with the superposition of two or three rather sharply peaked distributions, like the ones by Toufar \textit{et al.}~\cite{Toufar76}, Yu and Standish~\cite{Yu88}, or the various linear theories by De Larrard~\cite{DeLarrard99,Stovall86}. 
  Also for continuous size distributions a hierarchical partitioning model was recently developed~\cite{Kolonko10}.
  Most real grain-size distributions, used for dense packing, have a rather complicated shape often being a superposition of various empirical functions.
  It is therefore of interest to develop a technique to obtain an upper bound for a packing having arbitrary distribution.

  Based on the insight that a completely space-filling packing of spheres, i.e., having unity volume density, can only be achieved with a generalized (random) Apollonian setup~\cite{Oron00,Baram05}, we design a systematic technique to optimally fill the fines into the voids between larger grains and, by sweeping from the large end of any distribution, to finally obtain the highest density one could possibly attain with such distribution.
  Testing the technique on various real and artificial distributions we recover that power laws provide the highest densities.

  As illustrated in Fig.~\ref{fig::scheme}, we discretize the distribution into bins by grain size.
  Grains are assumed to have spherical shape (disks in two dimensions) and are organized in each bin in the monodisperse closest-packing configuration.
  Then the gaps are filled with smaller ones, following an Apollonian packing. 
  As rigorously proved by L. F. T\'oth \cite{Toth40,Toth60}, in two dimensions, the most efficient monodisperse arrangement of disks is the \textit{hexagonal closest packing} (\textit{hcp}), with a density of $\rho_\mathrm{hcp}=1/6\pi\sqrt{3}\approx0.9069$.
  With this arrangement, there are, per largest grain, two unitary voids to be filled with the traditional Apollonian packing (see Fig.~\ref{fig::scheme}).
  Each bin $b$ is characterized by four different parameters: the average radius, $r_b$; the volume of grains, $V_b$; the effective volume, $V_{\mathrm{eff}}^{b}$; and the density of the arrangement of particles in the bin, $\rho_b$.
  Since we consider all grains to have the same density the volume of grains, $V_b$, quantifies the total amount of matter.
  The excluded volume interaction between grains imposes geometrical restrictions such that, even for an efficient arrangement, the effective volume, $V_{\mathrm{eff}}^b$, is larger than $V_b$, accounting for the volume of both the grains and voids.
  The density of the arrangement in the bin, $\rho_b$, quantifies the efficiency of the packing and can be related with the volumes as $V_b=\rho_bV_{\mathrm{eff}}^b$.
 
  Figure~\ref{fig::scheme} is a pictorial scheme of the sequential compacting procedure, with bins representing grains grouped according to their size.
  The height of each bin represents the effective volume, $V_{\mathrm{eff}}^b$, and bins are organized in the inverse order of their radius.
  Initially, all bins are considered to be in the closest-packing configuration, \textit{hcp}, i.e., their density is $\rho_b=\rho_\mathrm{hcp}$.
  Starting with the bin $1$, corresponding to the largest grains, both the radius and volume of grains from further bins, required to fill the voids, are computed and transferred to the bin (details below).
  The volume of grains, $V_1$, and the density, $\rho_1$, are updated together with the effective volume and volume of grains of the bins from which particles are transferred.
  The process is executed for each bin, from the largest to the smallest grains, and the final net density, $\rho_\mathrm{net}$, is then computed as
    \begin{equation}\label{eq::netdensity}
      \rho_{net}= \frac{\sum_b V_b}{\sum_b V^b_\mathrm{eff}} \ \ ,
    \end{equation}
  where $V_b$ is the sum of the initial volume of particles in the bin with the volume of particles collected from the following ones.
  In the denominator, $V^b_\mathrm{eff}$ is the final effective volume of the bin.

  In the two dimensional case, for each generation of the Apollonian packing, the radius and the number of grains (disks) collected from other bins can be obtained, from the positive solutions of the Soddy-Gossett equation \cite{Lagarias02},
    \begin{equation}\label{eq::soddynD}
      2\left[\sum_{i=1}^{3}r_i^{-2}+r_j^{-2}\right]=\left[\sum_{i=1}^{3}r_i^{-1}+r_j^{-1}\right]^2 \ \ ,
    \end{equation}
  where the sum runs over the three disks limiting the void having radius, $r_i$, and $r_j$ is the radius of the required one. 
  The necessary volume of disks, needed to fill all voids, is then transferred from the corresponding bin. 
  If not enough volume is available, we proceed to the following bins until the required number of grains is collected.
\begin{figure}[b]
  \includegraphics[width=0.85\columnwidth]{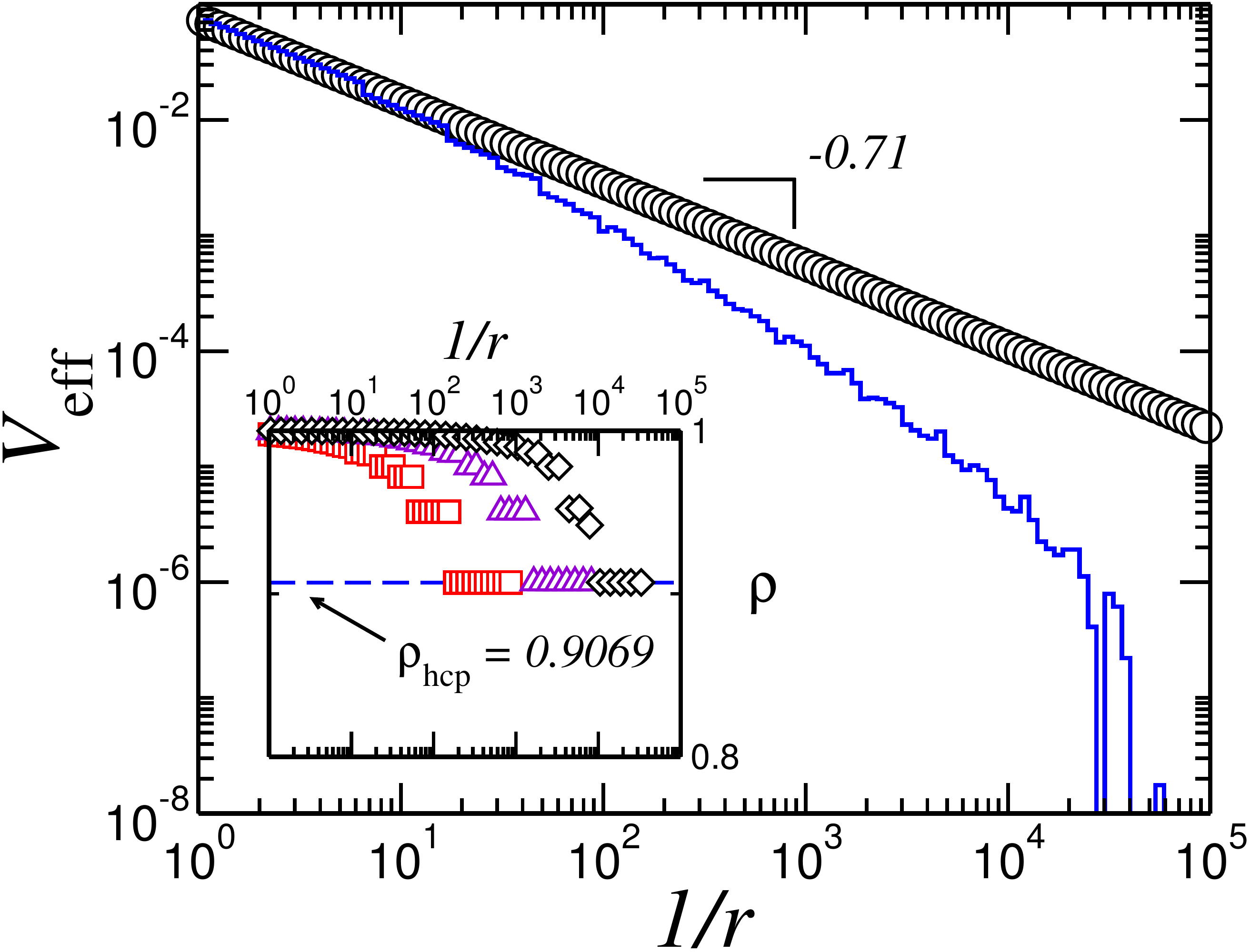}
  \caption{
    Initial and final distribution of the effective volume as a function of $1/r$, in 2d.
    The circles correspond to the initial distribution, given by $V_\mathrm{eff}\sim (1/r)^{-\alpha}$, with $\alpha=0.71$ and cutoff $\varepsilon=10^{-5}$ (minimum radius).
    The full-(blue) line, is the final distribution, with a density of $\rho_\mathrm{net}$.
    The inset shows the density of each bin as a function of $1/r$ (diamonds).
    Data for $\varepsilon=10^{-3}$ (squares) and $\varepsilon=10^{-4}$ (triangles) is also included.
    The dashed-(blue) line stands for the minimum density $\rho_\mathrm{hcp}\approx0.9069$, which corresponds to the initial configuration.
    \label{fig::effvol2d}
    }
\end{figure}
\begin{figure}
  \includegraphics[width=0.85\columnwidth]{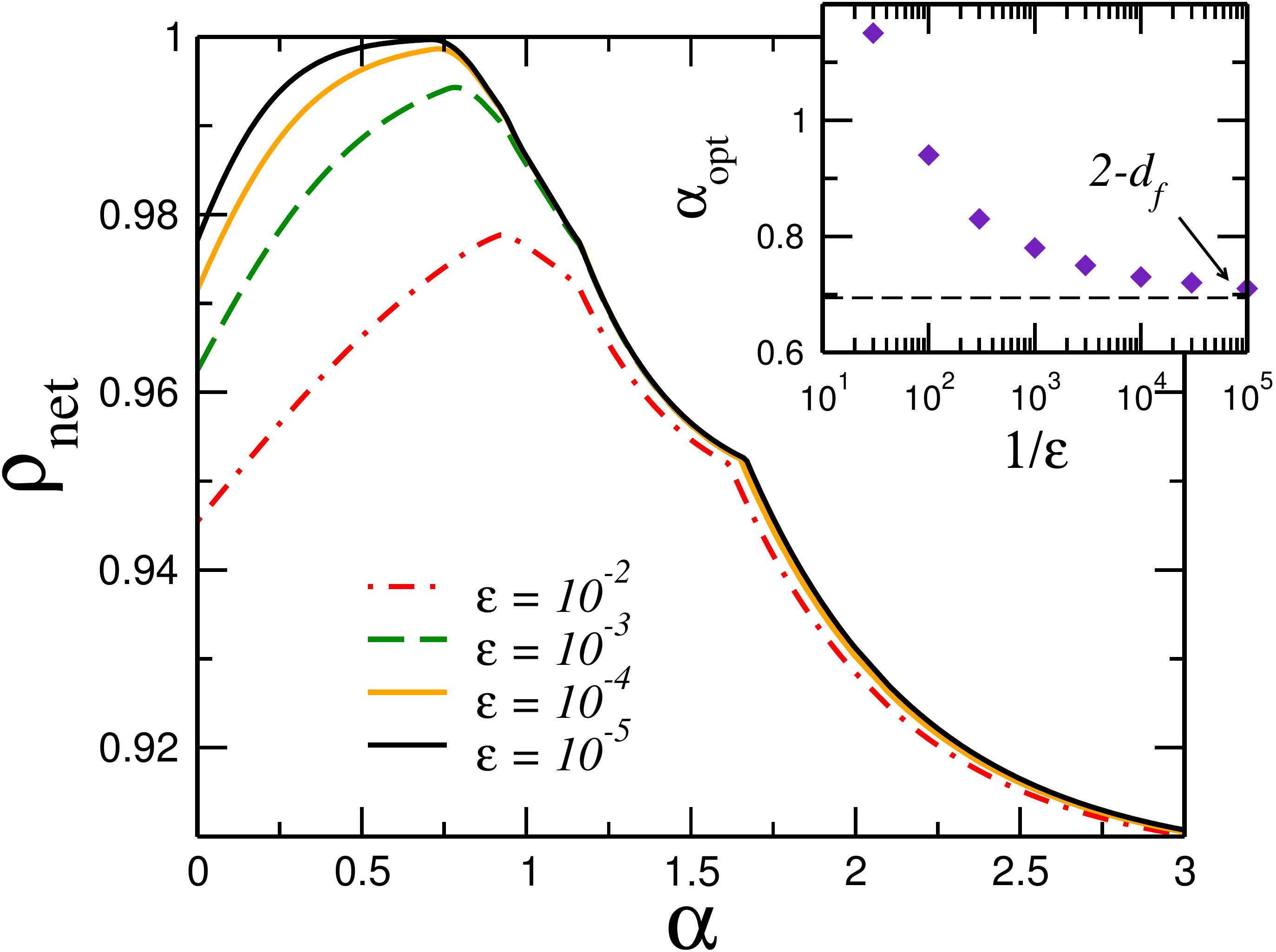}
  \caption{
    Final net density, $\rho_\mathrm{net}$, as a function of $\alpha$, for different values of the cutoff $\varepsilon$, in 2d.
    All curves are characterized by an optimal $\alpha=\alpha_\mathrm{opt}$, at which a maximum density is obtained.
    The inset shows the dependence of $\alpha_\mathrm{opt}$ on the inverse of the cutoff $\varepsilon$.
    When the effect of truncation is reduced the value of
$\alpha_\mathrm{opt}$ approaches $2-d_f$ (dashed line) as $\alpha_\mathrm{opt}-(2-d_f)\sim1/\varepsilon^{-a}$, where $d_f$ is the fractal dimension of the Apollonian packing and $a=0.37\pm0.09$.
    \label{fig::netdensity}
    }
\end{figure}

  Let us consider, as example, a grain-size distribution such that the volume of grains, $V_b$, is a power law of the grain size, i.e., $V_b\sim(1/r)^{-\alpha}$.
  As in real situations, the grain-size distribution is truncated at a lower cutoff, $\varepsilon$.
  Once initially grouped in bins, by size, and arranged in \textit{hcp}, $V_\mathrm{eff}\sim(1/r)^{-\alpha}$.
  Figure~\ref{fig::effvol2d} shows the initial (circles) and final (full line) effective volume, $V_\mathrm{eff}$, as a function of the inverse radius, for $\alpha=0.71$ and $\varepsilon=10^{-5}$.
  The proposed algorithm gives a final net density $\rho_\mathrm{net}\approx0.9997$, which is close to unity and a porosity $300$ times smaller then the closest-packing density of monodisperse disks.
  The effective volume of smaller-grain bins is significantly reduced and, for the tiny grains, the effective volume vanishes, i.e., all material is completely used to fill previous bins.
  The inset shows the final density of each bin.
  The density of the first bins is close to unity and the packing efficiency is solely limited by the cutoff.
  The smaller the grain size the lower the density since the number of Apollonian-packing generations which can be collected from further bins diminishes and the reserve from such bins also vanishes.
  The density of the final bins can either correspond to the one of the closest packing or be zero in the case that no material is left.
  In the same inset (Fig.~\ref{fig::effvol2d}) we also show the data for different cutoffs.
  For all considered values the same qualitative behavior is observed but the smaller the cutoff (minimum radius) the larger the number of different bins with density close to unity and the more effective the packing.

  Figure~\ref{fig::netdensity} shows the dependence of the final net density on the exponent $\alpha$, for different values of the cutoff, $\varepsilon$.
  For each cutoff, we observe an optimal value of $\alpha=\alpha_\mathrm{opt}$ at which a maximum final net density is obtained, that decreases with the cutoff, as shown in the inset of Fig.~\ref{fig::netdensity}.
  In the limit of vanishing cutoff, the optimal exponent converges to $2-d_f$, where $d_f$ is the fractal dimension of the Apollonian packing, which in 2d is $d_f\approx1.306$ \cite{Manna91}.
  In this optimal case, the density is unity and the distribution of grain volume is the one of the Apollonian packing \cite{Herrmann90,Manna91}.
  There exist different families of deterministic Apollonian packing with fractal dimensions ranging from $1.306$ to $1.802$ \cite{Oron00}, which will lead to different values $\alpha_\mathrm{opt}=d-d_f$ ($d$ is the dimension of the system).
  Random Apollonian packings, instead, are characterized by the same fractal dimension (as shown by Baram and Herrmann \cite{Baram05}) and have, therefore, the same $\alpha_\mathrm{opt}$.

  In 3d, grains are spheres and, for the sake of convenience, we start with a regular tetrahedron configuration with four mutually tangent spheres that are also tangent to an enveloping sphere, with a density $\rho\approx0.3633$ \cite{Borkovec94}.
  Recently, Baram and Herrmann \cite{Baram04b,Baram04} have introduced an algorithm to construct an Apollonian packing in 3d starting from any initial configuration, which we consider here to compute the radius and the number of grains to collect from further bins.
  Alike the 2d case, when an initial power-law distribution of volumes is considered, with $\alpha=0.55$ and cutoff $\varepsilon=3\times10^{-3}$, the final net density is $\rho\approx0.9325$, corresponding to a porosity that is $1/3$ of the one for the closest packing of monodisperse spheres.
  The $\alpha_\mathrm{opt}$ converges to $3-d_f$, with $d_f\approx2.4739$ the fractal dimension of the Apollonian packing \cite{Borkovec94}.
  For $\alpha=\alpha_\mathrm{opt}$, the porosity $p$ scales with the cutoff as $p\sim\varepsilon^\tau$, with $\tau=0.47\pm0.03$, close to the one of the Apollonian packing, which is $\tau=d-d_f\approx0.526$ \cite{Manna91,Herrmann90,Borkovec94}.
\begin{figure}[t]
  \includegraphics[width=0.85\columnwidth]{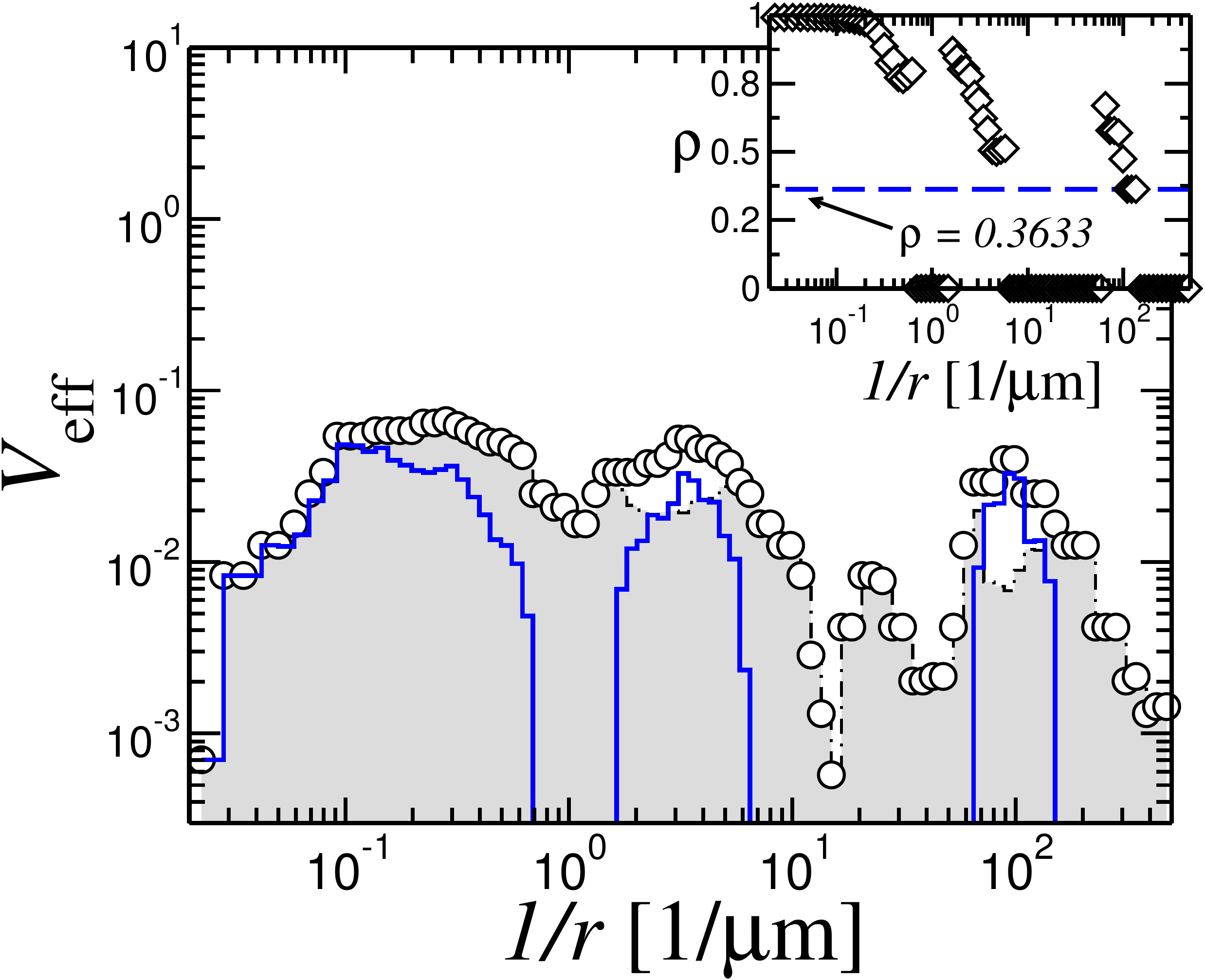}
  \caption{
    Initial and final distribution of the effective volume as a function of $1/r$, for 3d.
    The initial values were obtained from the empirical distribution of ultra-high performance concrete of Fig.~$7$ of Ref.~\cite{Geisenhansluke04}.
    The circles correspond to the initial distribution and the full-(blue) line is the final distribution, giving a density of $\rho_\mathrm{net}\approx0.8203$.
    The gray area stands for the optimal distribution proposed in the text, giving $\rho_\mathrm{opt}\approx0.9501$.
    As in Fig.~\ref{fig::effvol2d}, the inset shows the density of each bin as a function of $1/r$ (diamonds). 
    The dashed-(blue) line stands for the minimum density, which corresponds to the initial configuration.
    \label{fig::real}
  }
\end{figure}

\begin{figure}[t]
  \includegraphics[width=\columnwidth]{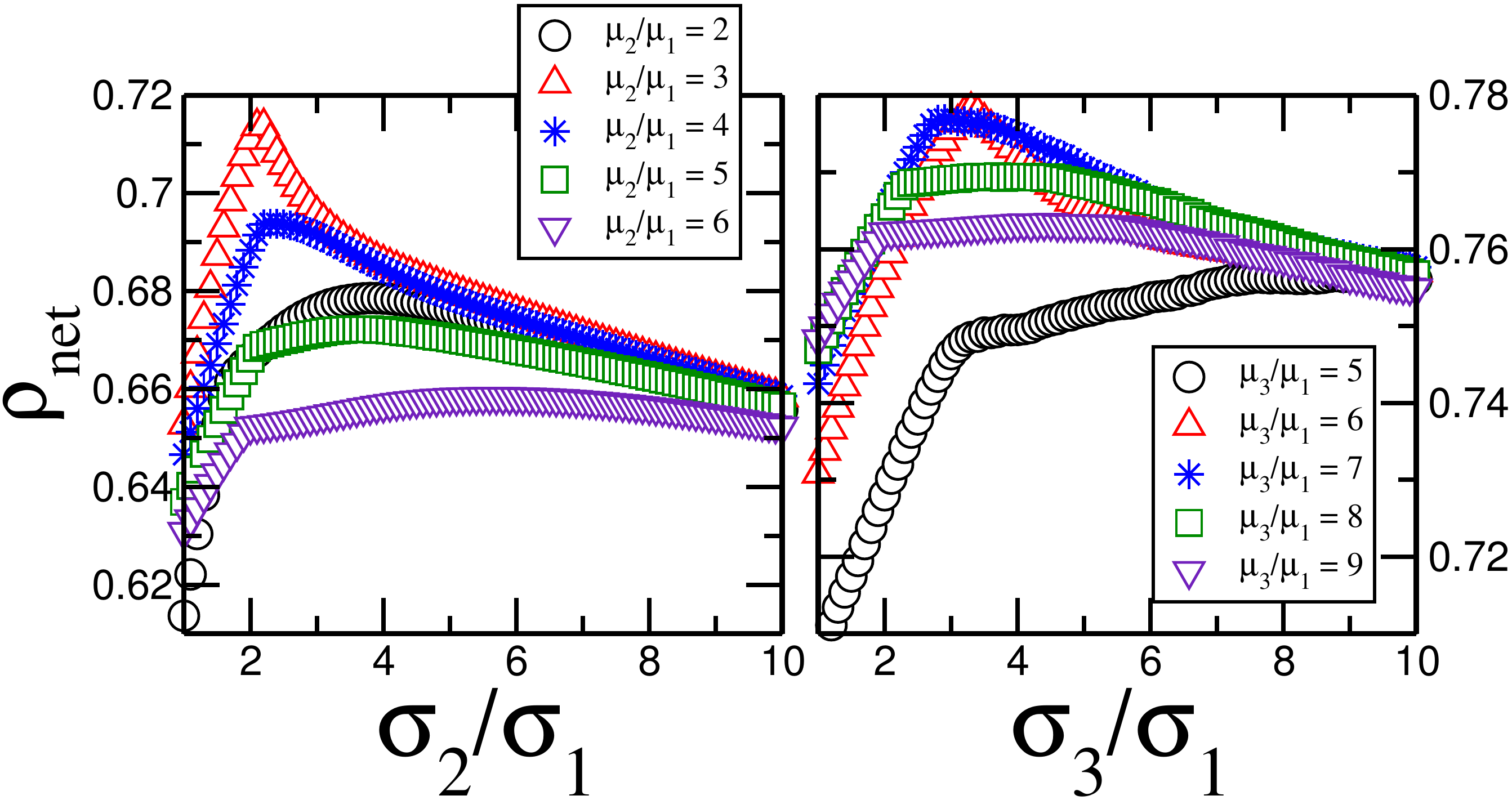}
  \caption{
    a) Final net density, $\rho_\mathrm{net}$, for two Gaussians as a function of the ratio between size dispersions ($\sigma_2/\sigma_1$). 
    Each set corresponds to different relative average sizes ($\mu_2/\mu_1$).
    b) Final net density for three Gaussians, where the first two have $\mu_2/\mu_1=3$ and $\sigma_2/\sigma_1=2.1$ (the highest density in a)), as a function of the size dispersion ($\sigma_3/\sigma_1$) and for different average sizes ($\mu_3/\mu_1$) of the third one.
    For both cases, $\varepsilon=3\times10^{-3}$.
    \label{fig::gaussians}
  }
\end{figure}

  In real applications, since a fine control of the grain size is not feasible, the usual distributions differ significantly from the idealized power laws.
  Due to the discrete nature of the system, the optimal distribution would be a sequence of delta functions centered at the characteristic grain sizes of the Apollonian arrangement and with prefactors given by a power law with exponent $\alpha_\mathrm{opt}$.
  But, such extremely narrow distributions are experimentally not realized and, instead, Gaussians with a certain size dispersion typically appear as the generic distribution in pratical situations and are used here to properly describe realistic grain-size distributions.
  Partitioners typically consider mixtures of materials with different characteristic grain sizes like, e.g., crushed quartz, cement, sand, silica fume, and microsilica, in the case of high performance concrete (HPC).
  To illustrate such procedure, in Fig.~\ref{fig::real} we take a representative empirical distribution for HPC, with a mixture of four components, obtained from Ref.~\cite{Geisenhansluke04}, for which we estimate an upper bound for the density of $\rho_\mathrm{net}\approx0.8203$ which is consistent with the typical values discussed in the reference (around $0.8$).
  Optimizing the grain-size distribution, by finding an efficient set of different sizes to minimize the porosity, is a relevant technological problem that can be addressed in a systematic way with our algorithm.
  For each component $i$, the distribution is characterized by a Gaussian with average size $\mu_i$, size dispersion $\sigma_i$, and height of the peak $H_i$.
  We start by considering two types of material, i.e., two Gaussians.
  For simplicity, to reduce the number of degrees of freedom we fix their volume fraction to be equal.
  By exploring the parameter space, the algorithm can identify the ratios between average sizes and size dispersions which minimize the porosity (see Fig.~\ref{fig::gaussians}a).
  Once this distribution has been identified, a third Gaussian can be included and its space of three parameters explored (see Fig.~\ref{fig::gaussians}b).
  The process can be repeated until the desired porosity is obtained.
  This gives an easy recipe to a systematic proportioning of mixtures for highly compacted powders.

  While in the absence of a cutoff the optimal distribution of grains is a power law with $\alpha=\alpha_\mathrm{opt}$, this is not the case for truncated ones.
  In these cases, the number of different Apollonian generations that can be collected from further bins is reduced as the average grain size of the bin decreases.
  When the hierarchical procedure is applied, not all material is used, as shown by the final distribution represented by the line in Fig.~\ref{fig::effvol2d}.
  The final net density can be increased if we subtract, from the initial distribution, the remaining material not considered in the arrangement.
  For example, we can subtract the volume of grains remaining inside all bins above the first one with vanishing $V_\mathrm{eff}$, which corresponds to the last grain size that is fully used to fill the voids between larger ones.
  Applying this strategy to the first two examples discussed in this Letter, we do not obtain for 2d (with $\alpha=0.71$ and $\varepsilon=10^{-5}$) any noticeable improvement, whereas in 3d (with $\alpha=0.55$ and $\varepsilon=3\times10^{-3}$), since the packing efficiency is significantly affected by these effects, $\rho_\mathrm{opt}\approx0.9493$.
  The latter corresponds to a decrease in almost $2\%$ in the porosity.
  The effect is even more significant with the empirical distribution in Fig.~\ref{fig::real}.
  In this case, a more efficient distribution is obtained (represented by the gray area in the plot), with a maximum density $\rho_\mathrm{opt}\approx0.9501$, which is three times lower in porosity than the original case.
  The proposed optimization technique could be implemented in practice for instance by using adequate filtering.

  In summary, we propose an iterative process to estimate the upper bound for the density of a packing of spheres having an arbitrary size distribution, which is meaningful for developing low-porosity materials.
  Grains are grouped by their sizes and, sweeping from the largest to the smallest grain size, the voids are hierarchically filled with smaller grains according to an Apollonian packing.
  We have analyzed the dependence of the optimal density on the properties of the distribution and recovered the power laws giving the most efficient case.
  We have suggested an iterative scheme to optimize the proportioning of the components in order to reach the lowest porosity.
  Future work should consider different shapes of grains or even a broad distribution of shapes.
  Developing experimental hierarchical procedures to obtain the efficient packings reported would be of paramount interest.

\acknowledgments
We acknowledge financial support from the ETH Risk Center.
We also acknowledge the Brazilian agencies CNPq, CAPES and FUNCAP, and the grant CNPq/FUNCAP, for financial support.

\bibliographystyle{eplbib}
\bibliography{packing}

\end{document}